\documentclass[twocolumn]{aastex62}

\submitjournal{ApJ}

\shorttitle{SN Ia UV Flux v. Metallicity Relation}
\shortauthors{Brown et al.}

\begin{document}

\title{A Photometric Analysis of the Relationship Between Type Ia Supernova \\ UV Flux and Host Galaxy Metallicity}

\correspondingauthor{Peter J. Brown}
\email{pbrown@physics.tamu.edu}
\author[0000-0001-6272-5507]{Peter J. Brown}
\affil{Department of Physics and Astronomy,
Texas A\&M University, 4242 TAMU, College Station, TX 77843, USA }
\affil{George P. and Cynthia Woods Mitchell Institute for Fundamental Physics \& Astronomy}

\author[0000-0002-8866-4797]{Nicole R. Crumpler}
\affil{Department of Physics and Astronomy,
Texas A\&M University, 4242 TAMU, College Station, TX 77843, USA }
\affiliation{Department of Physics, Duke University, Durham, NC 27708, USA}

\begin{abstract}

The effect of progenitor metallicity on Type Ia Supernovae (SNe Ia) 
has important cosmological implications due to the need for these standardizable candles to be compared across large spans of cosmic time in which the progenitor stars might have different properties.  
Theoretical models have come to different conclusions as to the wavelength range impacted by metallicity differences, leading to differing interpretations of the growing sample of UV observations.  Recent work has claimed a correlation between the mid-UV flux of SNe Ia measured from Swift grism spectra and the gas-phase metallicities measured for their host galaxies.  
Here we examine UV photometry for the same objects.  We find no significant correlations between the UV-optical colors (or UV/optical count rate ratios) of the SNe Ia and the host galaxy properties of mass or metallicity.  The lack of a significant correlation with host galaxy metallicity implies another physical difference other than progenitor metallicity dominates the UV flux differences.  
We are no longer limited by a lack of UV observations.  Rather, understanding the existing observations requires improved theoretical models and a larger parameter space of physical differences.

\end{abstract}


\section{Introduction} \label{sec:intro}

Type Ia Supernovae (SNe Ia) are important cosmological tools because their similarly high luminosities allow them 
to be used as standard candles to measure distances.  
The dispersion in luminosities can be reduced with corrections for the light curve shape  and colors \citep{Phillips_etal_1999,Riess_etal_1996_mlcs,Hamuy_etal_1996H0}.  
Correlations have also been found with host galaxy mass \citep{Kelly_etal_2010, Lampeitl_etal_2010,Childress_etal_2013, Hayden_etal_2013,Moreno_etal_2016} and are now corrected for in most cosmological studies \citep{Scolnic_etal_2018}.  A variation in the magnitude of the so-called ``mass step'' have been found for different surveys, filter choices, and SN fitters \citep{Kim_etal_2014, Brout_etal_2019}.  Correlations have also been found with the local host galaxy parameters as well \citep{Rigault_etal_2013,Galbany_etal_2014}.  The physical connection between the host galaxy parameters and the SN luminosities is not yet clear.  
The concern is that systematic differences in the progenitors could change the observed properties of the explosion in a way not corrected for 
in the current methods of standardization.  

Metallicity, here referring to the pre-explosion chemical abundances in the white dwarf, has been worried about as a potential systematic for many years \citep{Hoeflich_etal_1998, Mazzali_Podsiadlowski_2006} because of the secular building up of metals in the universe and the different chemical compositions of the galaxies hosting SNe Ia.
Many models have explored how changes in progenitor metallicity will effect the bolometric and broadband fluxes of SNe Ia (\citealp{Timmes_etal_2003}, see also  \citealp{De_etal_2014} and references therein).  Those which calculated effects at UV wavelengths found stronger effects in the UV than in the optical \citep{Lentz_etal_2000,Sauer_etal_2008,Walker_etal_2012,Miles_etal_2016}, though differing from each other in how much or in what wavelength regions the effects are seen.


\citet{Foley_Kirshner_2013} presented UV spectra of two SNe Ia which are nearly identical twins in terms of near-UV/optical spectra and light curve shapes.  The most significant difference is in the mid-UV flux level.  By comparison with the models of \citet{Lentz_etal_2000}, they found a relative metallicity difference could approximate the spectral changes in the mid-UV.  \citet{Brown_etal_2015} found that the same metallicity differences could be inferred from Swift/UVOT broad-band photometry given the assumption that the color variations were caused by the same metallicity differences as between the epoch 15 \citet{Lentz_etal_2000} models.  \citet{Brown_etal_2015} also showed, however, that the UV flux levels of the \citet{Lentz_etal_2000} models were too high and that even the relative color differences between objects did not match the multi-epoch color differences of SNe 2011by and 2011fe.  
Using a different set of models from \citet{Walker_etal_2012}, \citet{Brown_etal_2017erp} found that SN~2011fe had UV flux levels comparable to low-metallicity models while the redder SN~2017erp featured a depressed UV continuum which matched the higher-metallicity models.  From the observations, it is clear that significant but distinct variations appear in the mid-UV and near-UV wavelength regions; however, the same physical difference -- metallicity -- has been invoked for both of the spectral ranges.

A promising approach is to look for other observables which correlate with the UV differences (e.g. \citealp{Milne_etal_2013,Brown_etal_2018}).  \citet{Pan_etal_2019} report a correlation between the host galaxy metallicity and f$_{2535}$,the ratio of the mid-UV flux between 2450-2620 \AA~and the optical flux between 4000-4500 \AA, as measured in grism observations with the Neil Gehrels Swift Observatory's Ultra-Violet Optical Telescope \citep{Gehrels_etal_2004, Roming_etal_2005,Kuin_etal_2015}.  In this article we report whether we can confirm such a correlation from the Swift/UVOT photometry of the same objects.  A shown below, the photometry does not show a correlation with the host galaxy properties.

\section{Observations} \label{sec_obs}

The photometry used is from the Swift Optical Ultraviolet Supernova Archive (SOUSA; \citealp{Brown_etal_2014_SOUSA}).
Most of the photometry has been previously published and made available via the Swift SN website
and the Open Supernova Catalogue (OSC; \citealp{Guillochon_etal_2017}), while four additional objects were reduced and added following the posting of \citet{Pan_etal_2019}.  

Some objects from \citet{Pan_etal_2019} are not included, as we do not yet have pre-explosion or late-time images from which to subtract the underlying galaxy contribution to the observed count rates.  This turns out to be especially important for the sample used by  \citet{Pan_etal_2019}.  SN~2016ccz has the lowest host-galaxy metallicity in the \citet{Pan_etal_2019} sample and is claimed to have the highest mid-UV/optical flux ratio.  This object dominates the UV-metallicity relationship in \citet{Pan_etal_2019}.  In our photometry analysis, we do not significantly detect SN2016ccz in the mid-UV uvm2 filter after subtraction of the host galaxy flux. While the \citet{Pan_etal_2018} UVOT grism reduction has some improvements compared to the default extractions, it can only estimate the underlying contribution based on the background flux on either side of the source spectrum.  It does not use the more accurate spectral template image subtraction as in \citet{Smitka_etal_2016}. 
This is necessary in cases of strong, non-uniform galaxy contamination.  In the case of SN2016ccz, the mid-UV flux in their spectrum must be dominated by the galaxy.  As SN2016ccz is not detected significantly and has a negative count rate consistent with zero flux, it would be excluded or treated as an upper limit if fitting the magnitudes or colors.  Instead, we use the count rates and count rate errors at the epoch closest to the time of B-band maximum light to test for correlations with the host galaxy mass or metallicity.

We use the host galaxy parameters as measured by \citet{Pan_etal_2019}.  They  derived host galaxy masses from fits to multicolor photometry.  The host galaxy metallicities are calculated from emission line measurements from optical spectra on the \citet{Pettini_Pagel_2004} calibration.

\section{Analysis} \label{sec_analysis}

We first confirm that the observed colors are similar to those of other UV photometric Ia samples studied previously (e.g. \citealp{Brown_etal_2010,Milne_etal_2010,Milne_etal_2013,Brown_etal_2017}).  Figure 1 shows the SN Ia color evolution plotted with color regions signifying NUV-blue and NUV-red SNe as designated by \citet{Milne_etal_2013}. By fraction, the NUV-red SNe have previously been observed to be more common \citep{Milne_etal_2013,Milne_etal_2015} at low redshift.  This sample seems to have a larger fraction of NUV-blue SNe Ia.  The spectroscopic nature of this sample could be biased, either in triggering or in publishing SNe Ia with higher (and thus more detectable) UV flux.  The higher flux could result from intrinsic differences or from lower host galaxy or MW reddening. However, a similar range of colors is observed. 

\begin{figure*}[htbp]
\begin{center}
\includegraphics[scale=0.85,angle=0]{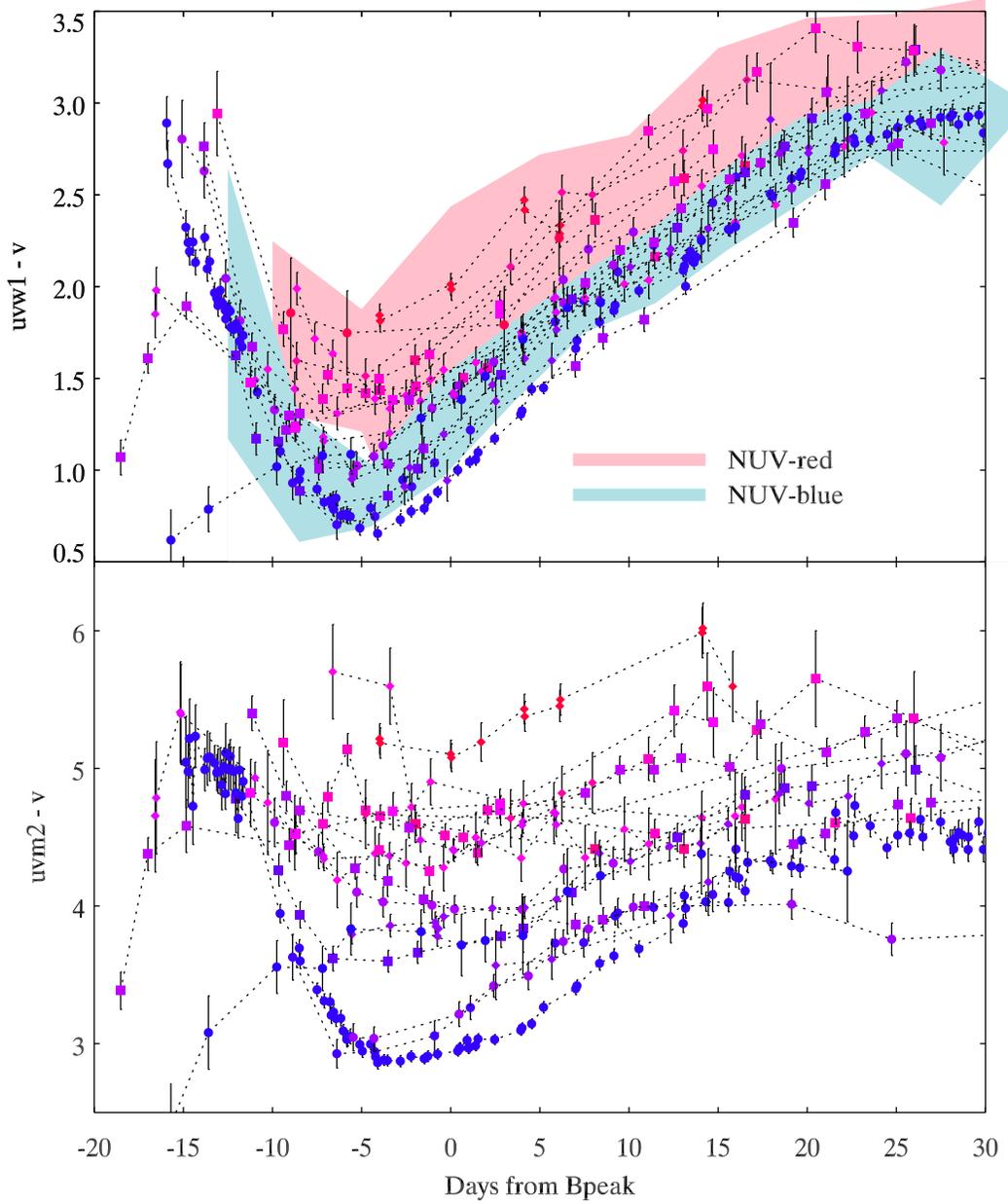}
\caption{uvw1-v and uvm2-uvw1 color evolution for the SNe Ia.  This sample is identical to \citet{Pan_etal_2019} except for the exclusion of the rapidly-declining SNe 2005ke and 2011iv and several SNe Ia which lack template images for host galaxy flux subtraction. The colors of these SNe Ia cover a similar range as those studied previously by \citet{Milne_etal_2010} and \citet{Brown_etal_2017}.  The Swift/UVOT photometry gives a better view of the relative flux evolution than from the limited spectra available in \citet{Pan_etal_2019}.  Earlier than 5 days before B-band maximum light it is still hard to judge the spread in colors due to the faintness of most the objects in the mid-UV.  A few objects, notably iPTF14bdn \citep{Smitka_etal_2015} and SN~2017cbv \citep{Hosseinzadeh_etal_2017cbv} were bluer at earlier times than the majority of SNe Ia. \label{fig_colorevolution}}
\end{center}
\end{figure*}

\begin{figure*}[!tbp]
  \centering
  \begin{minipage}[b]{0.45\textwidth}
    \includegraphics[width=\textwidth]{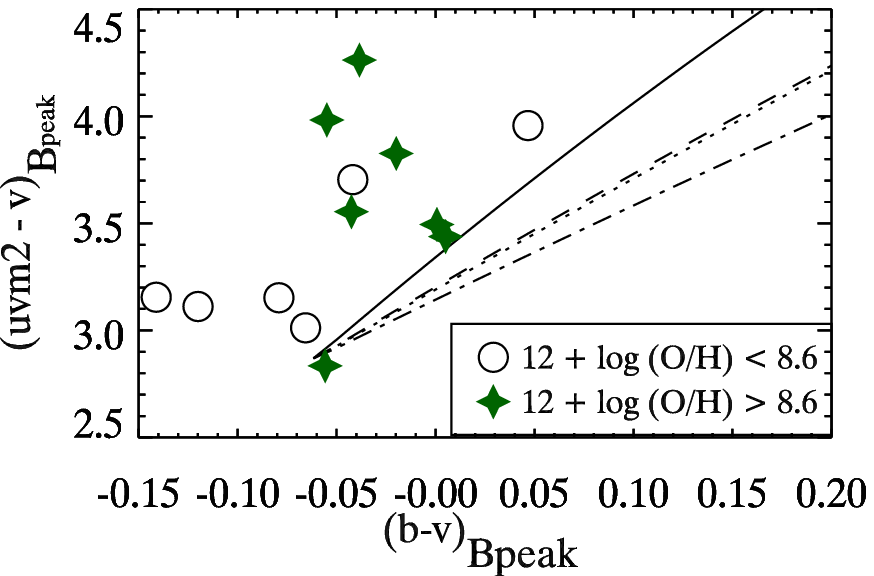}
  \end{minipage}
  \hfill
  \begin{minipage}[b]{0.45\textwidth}
    \includegraphics[width=\textwidth]{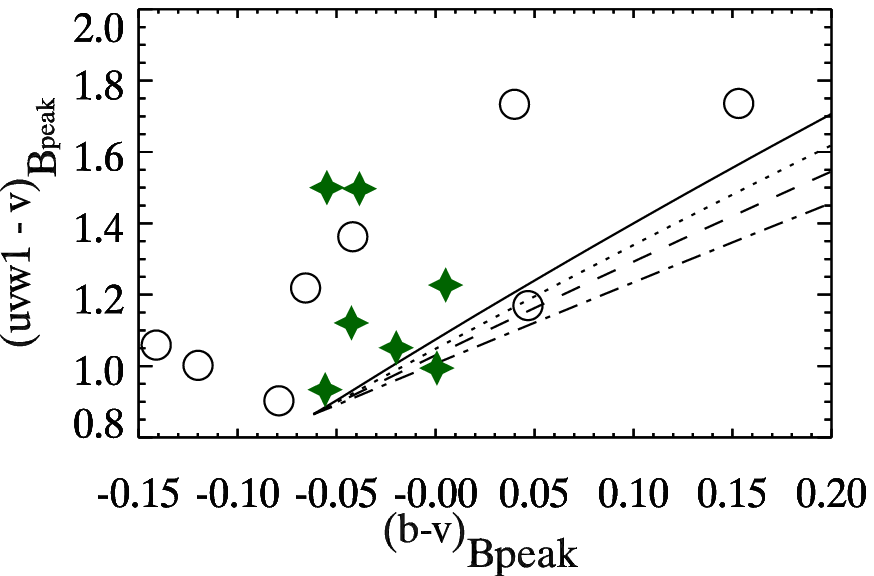}
  \end{minipage}
  \caption{UV colors versus optical colors observed at the time of maximum light in the B band.  The lines represent the SN~2011fe spectrum reddened by different extinction laws (see \citealp{Brown_etal_2017}).  The intrinsic dispersion is apparent by the vertical spread in points which is present for SNe Ia from low and high metallicity host galaxies.}
  \label{fig_peakcolors}
\end{figure*}

The color evolution plots are not corrected for reddening, which can have the effect of shifting a NUV-blue SN Ia into the NUV-red parameter space \citep{Brown_etal_2017}.  To account for reddening, Figure \ref{fig_peakcolors} displays the observed colors, but with a reddening track showing the effect of different reddening laws.  The vertical dispersion is indicative of intrinsic UV differences.  Splitting the sample at 12 + log(OH)=8.6, we find that both groups disperse vertically, though the higher metallicity group has a smaller spread in the $B-V$ color.  Regardless of whether the NUV-blue/red groups are distinct or a continuous color parameter, they are not related to the host galaxy metallicity.  This seems to contradict the conclusions of \citet{Brown_etal_2017erp} if the metallicity differences of the \citet{Walker_etal_2012} models are attributed to the progenitor metallicity.  Recent measurements of the gas phase metallicity in two star forming regions in NGC~5861 point to a subsolar metallicity (12 + log(OH)=7.67 $\pm$ 0.01 and 7.87 $\pm$ 0.03; \citealp{Munoz_etal_2019}) for the host of the NUV-red SN~2017erp.

We now look directly for correlations between the UV flux of this SN Ia sample and the properties of the host galaxies.
To make a clearer comparison with the UV-optical flux ratios shown in \citet{Pan_etal_2019}, we use the UVOT filter uvm2 for the mid-UV, and uvw1 and u for the near-UV, and B for the optical (sometimes designated as $b$ as per the UVOT convention).  To incorporate the undetected flux of SN~2016ccz we now use measured count rates and associated errors at the epoch closest to maximum light in the B band rather than the fitted magnitudes.  We first correct the count rates for the estimated Milky Way reddening \citep{Schlafly_Finkbeiner_2011} and the host galaxy reddening using the values from \citet{Pan_etal_2019}.  Some of the \citet{Pan_etal_2019} host galaxy reddening values are actually from the galaxy spectral fitting and thus not necessarily reflective of the line of sight reddening to a SN in or near a particular part of the host galaxy.  We do a separate correction assuming that all of the SNe Ia have the same $B-V$ color at the time of maximum in the B band, set to the bluest object in this sample with (B-V)$_{Bpeak}$=-0.15.  The count rates are corrected for reddening using the extinction coefficients derived from the \citet{Cardelli_etal_1989} reddening law with R$_V$=3.1 and the SN~2011fe spectrum \citep{Brown_etal_2016}.

We can confirm a trend of colors with the light curve shape (stretch or $\Delta $M$_{15}$(B), as also seen in \citealt{Brown_etal_2010} and \citealt{Foley_etal_2016}), so we exclude the rapidly-declining SNe 2005ke and 2011iv from this analysis to focus on the UV differences of SNe Ia with more normal light curve shapes.

\begin{figure*}[h!]
\begin{center}
\includegraphics[scale=0.85,angle=0]{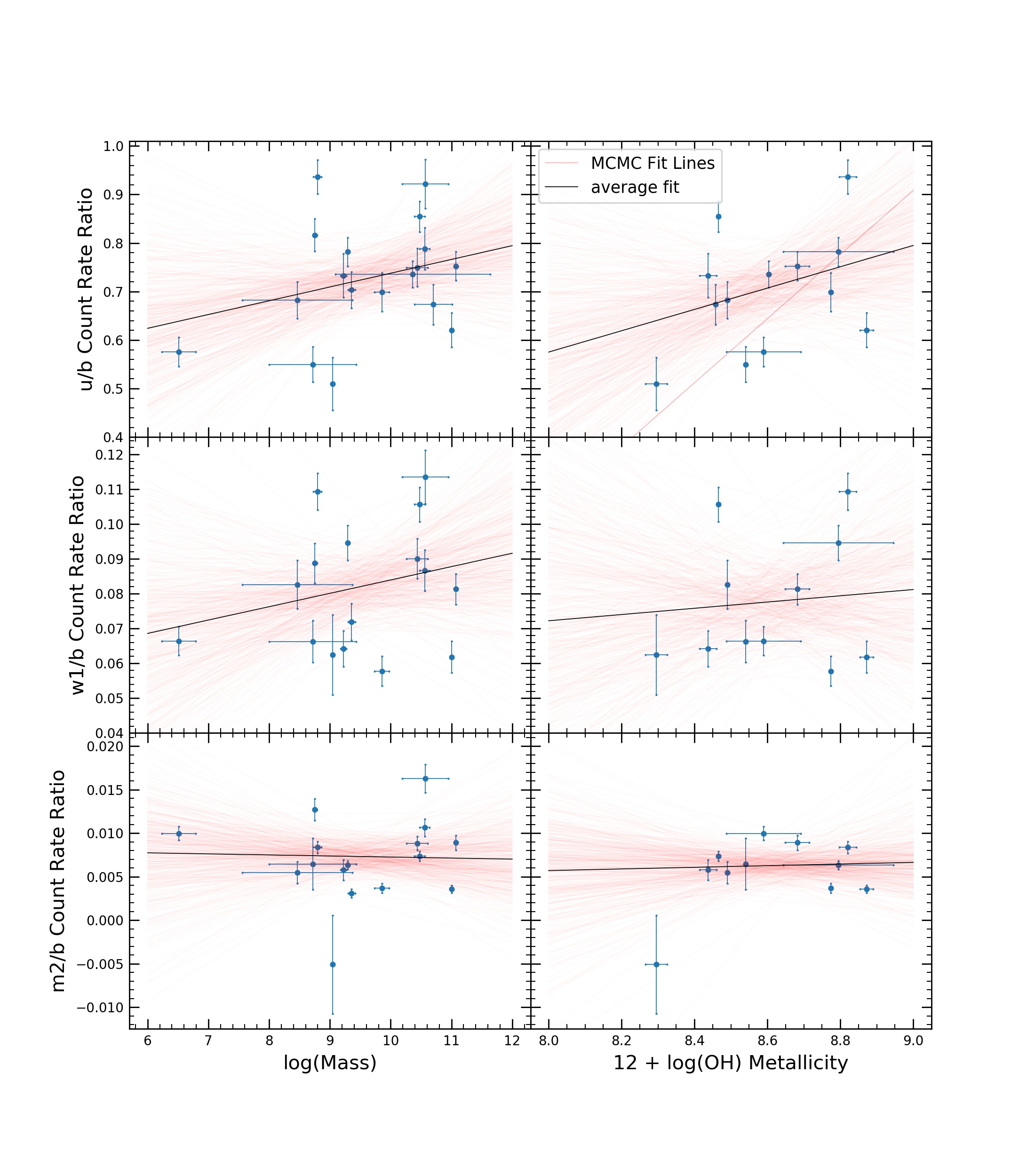}

\caption{UV/optical count rate ratios of the SNe Ia comparied to the mass and metallicity of the host galaxy. Based on the MCMC fit parameters displayed in Table 1, we do not find sufficient evidence for a correlation between the mass or metallicity of SN Ia host galaxies and the UV/optical count rate ratios from Swift/UVOT. \label{fig_massmetalicity}}
\end{center}
\end{figure*}

Figure \ref{fig_massmetalicity} shows the count rate ratios (with the color-based reddening correction) plotted with respect to the host galaxy mass and metallicity as reported by \citet{Pan_etal_2019}.
Linear fits are performed using the Python Markov chain Monte Carlo (MCMC) Bayesian linear fitting algorithm LINMIX \citep{Kelly_2007}. This process is able to account for heteroscedastic errors in both the x and y components of the fit.  The uncertainties in the host galaxy parameters from \citet{Pan_etal_2019} are not symmetric.  In Table 1 we report results using the positive error bars for both, though we have checked that our conclusions are unchanged if we use just the negative error bars.  We execute the MCMC fit with 10,000 draws from the posterior distribution and K=2 Gaussians in the mixture model. We provide the mean slope and intercept from the fits as well as the standard deviation of both of these parameters in Table 1. Values are provided for our color-based extinction correction as well as the original \citet{Pan_etal_2019} host reddening estimates.   The mean slopes show a different trend than that seen by \citet{Pan_etal_2019}, namely we see stronger trends in the near-UV u and uvw1 filters than in the mid-UV uvm2 filter.  Nevertheless, a significant fraction of the draws from the posterior distribution have positive and negative slopes, and the mean slope is never greater than one standard deviation of the sampled slopes. The limited number of SNe Ia with low host masses and low host galaxy metallicities gives individual objects a disproportionate weight considering the large scatter in the more population regions of the host galaxy parameter space.

We also present the Kendall and Pearson correlation coefficients and p-values. These do not take into account the uncertainties, but are independent means of evaluating the significance of a correlation.  The p-values are high, with the smallest being 0.16. 
Thus we do not find evidence with which to reject the null hypothesis that there is no correlation between the host galaxy mass or metallicity and the UV-optical count rate ratios at the time of B-band maximum light.

\section{Discussion}

The lack of correlations between the UV flux and the host galaxy metallicity is surprising given the many models which predict strong UV effects. 

There are a number of reasons which might contribute to a lack of correlation between the UV properties of the SNe Ia and the measured metallicities of their host galaxies.
First, within a galaxy there are metallicity gradients \citep{Zaritsky_etal_1994}, such that the global gas-phase metallicity may not represent the gas-phase metallicity at the location of the SN progenitor.
\citet{Galbany_etal_2016} found that on average there are small differences between the total and local metallicity.  These differences were measured as -0.018 and -0.088 dex, respectively, for the oxygen and stellar metallicities.  For the oxygen abundances, the metallicity measured at the location of individual objects changed from 0.1 below to 0.1 above that of the host galaxy as a whole, while the stellar metallicity differences ranged from -~0.3 to 0.2 dex. 
\citet{Zhou_etal_2019} used integral field spectroscopy to study the local, central, and global metallicities of SN hosts.  They conclude that the bias is small between them, though individual objects varied by up to 0.6 dex.  Though small, these metallicity differences within a galaxy are as large as the range of this \citet{Pan_etal_2019} sample.  
To give an example noted by \citet{Foley_Kirshner_2013}, the SDSS-measured metallicity of M101 is 9.12 \citep{Prieto_etal_2008}, but the gradient results in a value of 8.45 at the location of SN~2011fe (\citealp{Stoll_etal_2011}, and see also \citealp{Vilchez_etal_2019}).  
The global versus local metallicity difference for SN~2011fe is as large as that between the lowest and highest metallicities of the galaxies in \citet{Pan_etal_2019} sample.  The lack of a correlation with the {\it global} host galaxy metallicity does not rule out the possibility of a correlation with the {\it metallicity at the SN location}.

Second, the time between the birth of a star and when it eventually explodes  leads to a difference between the progenitor metallicity and the gas-phase metallicity measured near the time and place of explosion due to the chemical evolution of the galaxy \citep{Wyse_Silk_1987} and the migration of the progenitor from its birth place.  This depends on the delay time distribution between progenitor formation and explosion (see \citealp{Bravo_Badenes_2011} and the discussion therein).  SNe Ia in passive galaxies are found to have a larger difference than those in actively star-forming galaxies.  If SNe Ia have the same birth to explosion delay and the underlying host galaxies evolve in a uniform matter, the metallicities would be uniformly shifted.  Differences in the delay time and evolution of the galaxy metallicities, however, would lead to more scatter.  The lack of a correlation with the {\it current} host galaxy metallicity does not rule out the possibility of a correlation with the {\it progenitor metallicity at the time the progenitor formed}.

Third, the white dwarf progenitor may have a different composition at the time of explosion than it did at birth.  \citet{De_etal_2014} note that simmering causes a neutronization floor larger than that from the progenitor metallicity.  They highlight that the most relevant parameter is the metallicity at the time of explosion.  The lack of a correlation with the host galaxy metallicity does not rule out the possibility of a correlation with the {\it SN progenitor metallicity at the time of explosion}.

Finally, the explosion itself results in an mixing of the products of nucleosynthesis which can mimic the effect of metallicity differences \citep{Hoeflich_etal_1998}.
The models of \citet{Walker_etal_2012} acknowledge that the adopted composition structure could be affected by upmixing as well as primordial metallicity. The lack of an observed correlation with the host galaxy metallicity does not rule out the possibility of a correlation with the  {\it metal content in the SN ejecta}.

There are also complications of the observations and in the analysis which could mask a real correlation.  
While the UV photometry used here has much higher S/N than the spectra used by \citet{Pan_etal_2019}, we have a comparable (and slightly smaller) sample.  This is particularly problematic at the low mass and low metallicity regions where single objects have an enhanced impact on the correlations.  
Even with this sample, however, it is clear that there is significant scatter beyond any possible correlation.  Such a scatter points to other physics which is at least as important as progenitor metallicity to the formation and absorption of the UV flux.  The next advances will require going beyond the existing models.  Self-consistent modeling of the progenitor evolution, nucleosynthesis, radiative transfer, and computation of observables is needed \citep{Bravo_etal_2010,Miles_etal_2016}.

A further complication in UV analyses is the correction for dust reddening.  In particular, if a relevant variable such as metallicity also affects the optical colors from which the reddening is estimated \citep{Hoeflich_etal_2000,Walker_etal_2012,Brown_etal_2017erp}, an inaccurate correction of the UV flux could enhance or diminish the physical effect being studied.   \citet{Brown_etal_2018} showed this effect in a study of the UV colors of SNe Ia of differing velocities for which the optical color effect is still being studied.
This is important in studies such as this, as the amount and type of dust may also correlate with the host galaxy properties.

\section{Conclusions} \label{sec_conclusions}

In conclusion, we find no evidence for a correlation between the mass or metallicity and the UV-optical colors or count rate ratios.  This in contrast to the findings of \citet{Pan_etal_2019} which we attribute to an overestimate of the UV flux for SN~2016ccz  in the lowest metallicity host galaxy of the sample.  Samples covering a wider range in host galaxy parameters will help elucidate any physical connections which may exist.  The large scatter seen perpendicular to correlations of the UV flux with light curve shape \citep{Brown_etal_2010,Foley_etal_2016}, optical colors (as a proxy for reddening; \citealp{Brown_etal_2016}), velocity \citep{Milne_etal_2013,Brown_etal_2018}, and now host galaxy mass and metallicity imply yet another physical difference must exist between optically-similar SNe Ia which strongly affects the UV.  Further theoretical work is needed to compare against the multiwavelength and time-series data now available to better understand the diversity of these standard candles and how it might impact their use as cosmological distance indicators.

\acknowledgments

PJB receives support through program \#14665 provided by NASA through a grant from the Space Telescope Science Institute, which is operated by the Association of Universities for Research in Astronomy, Inc., under NASA contract NAS 5-26555.  
PJB is also supported by the Mitchell Institute for Fundamental Physics and Astronomy at Texas A\&M University.
This work was supported by NSF grant AST-1560223 (PI: Marshall) ``REU Site: Astronomical Research and Instrumentation at Texas A\&M University.'' We appreciate helpful comments from E. Baron, X. Yan, and N. Suntzeff.

\vspace{5mm}
\facilities{Swift(UVOT)}

\software{HEASOFT, CALDB,  
IDL, astrolib,  python, astropy, scipy, linmix }



\bibliographystyle{aasjournal}

\begin{deluxetable}{llrrrrrrrc}\label{table_fits}
\tablecaption{MCMC Linear Fit Parameters}
\tablehead{ \colhead{Host Parameter} & \colhead{Reddening} & \colhead{Color} & \colhead{Slope} & \colhead{Intercept} & \colhead{Kendall} & \colhead{Kendall} & \colhead{Pearson} & \colhead{Pearson} & \colhead{\% of Fits} \\ 
\colhead{ } & \colhead{Correction} & \colhead{Ratio} & \colhead{} & \colhead{} & \colhead{Correlation} & \colhead{P-Value} & \colhead{Correlation} & \colhead{P-Value} & \colhead{with Negative} \\ 
\colhead{} & \colhead{} & \colhead{} & \colhead{} & \colhead{} & \colhead{Coefficient} & \colhead{} & \colhead{Coefficient} & \colhead{} & \colhead{Slope} \\ 
}
\startdata
log (M) & P19 & u/b & 0.03$\pm$0.03 & 0.39$\pm$0.28 & 0.15 & 0.44 & 0.31 & 0.23 & 14 \\
log (M) & P19 & w1/b & 0.00$\pm$0.01 & 0.05$\pm$0.05 & 0.16 & 0.44 & 0.19 & 0.50 & 36 \\
log (M) & P19 & m2/b & -0.00$\pm$0.00 & 0.01$\pm$0.01 & -0.11 & 0.63 & -0.05 & 0.87 & 70 \\
12 + log (O/H) & P19 & w1/b & 0.02$\pm$0.04 & -0.11$\pm$0.37 & 0.20 & 0.45 & 0.25 & 0.47 & 31 \\
12 + log (O/H) & P19 & m2/b & 0.00$\pm$0.01 & -0.01$\pm$0.04 & 0.16 & 0.54 & 0.39 & 0.24 & 36 \\
log (M) & color & u/b & 0.03$\pm$0.03 & 0.45$\pm$0.30 & 0.18 & 0.35 & 0.32 & 0.21 & 15 \\
log (M) & color & w1/b & 0.00$\pm$0.01 & 0.05$\pm$0.05 & 0.11 & 0.63 & 0.26 & 0.36 & 21 \\
log (M) & color & m2/b & -0.0$\pm$0.00 & 0.01$\pm$0.01 & 0.05 & 0.85 & 0.05 & 0.86 & 58 \\
12 + log (O/H) & color & u/b & 0.22$\pm$0.24 & -1.18$\pm$2.08 & 0.31 & 0.16 & 0.39 & 0.19 & 17 \\
12 + log (O/H) & color & w1/b & 0.01$\pm$0.05 & 0.00$\pm$0.39 & 0.13 & 0.65 & 0.18 & 0.60 & 42 \\
12 + log (O/H) & color & m2/b & 0.00$\pm$0.01 & -0.00$\pm$0.07 & 0.09 & 0.76 & 0.41 & 0.22 & 51 \\
\enddata
\end{deluxetable}

\end{document}